\documentclass[a4paper,11pt]{article}
\usepackage{arxiv}
\usepackage{amsmath,amsfonts,amssymb}
\usepackage{graphicx}
\usepackage[colorlinks=true, allcolors=blue]{hyperref}
\usepackage{threeparttable} 
\usepackage{macro_ref}
\usepackage{geometry}
\usepackage{natbib}

\title{Pushing high angular resolution and high contrast observations on the VLTI from Y to L band with the Asgard instrumental suite: integration status and plans}

\renewcommand{\undertitle}{}
\renewcommand{\headeright}{}
\renewcommand{\shorttitle}{Asgard}
\date{}

\begin{document} 
\maketitle

\begin{center}
\vspace{-5em}
Marc-Antoine Martinod$^a$\footnote{marc-antoine.martinod@kuleuven.be},
Denis Defrère$^a$,
Michael J. Ireland$^b$,
Stefan Kraus$^c$,
Frantz Martinache$^d$,
Peter G. Tuthill$^e$,
Fatmé Allouche$^d$,
Emilie Bouzerand$^g$,
Julia Bryant$^e$,
Josh Carter$^b$,
Sorabh Chhabra$^c$,
Benjamin Courtney-Barrer$^b$,
Fred Crous$^e$,
Nick Cvetojevic$^d$,
Colin Dandumont$^f$,
Steve Ertel$^{h,i}$,
Tyler Gardner$^c$,
Germain Garreau$^a$,
Adrian M. Glauser$^g$,
Xavier Haubois$^k$,
Lucas Labadie$^j$,
Stéphane Lagarde$^d$,
Daniel Lancaster$^c$,
Romain Laugier$^a$,
Alexandra Mazzoli$^f$,
Anthony Meilland$^d$,
Kwinten Missiaen$^a$,
Sébastien Morel$^d$,
Daniel J. Mortimer$^l$,
Barnaby Norris$^e$,
Jyotirmay Paul$^c$,
Gert Raskin$^a$,
Sylvie Robbe-Dubois$^d$,
J. Gordon Robertson$^e$,
Ahmed Sanny$^j$,
Nicolas Schuhler$^k$,
Owain Snaith$^c$,
Adam Taras$^e$.
~\newline
~\newline
$^{a}$Institute of Astronomy, KU Leuven, Celestijnenlaan 200D, 3001 Leuven, Belgium;
$^{b}$Research School of Astronomy and Astrophysics, College of Science, Australian National University, Canberra 2611, Australia;
$^{c}$School of Physics and Astronomy, University of Exeter, Stocker Road, Exeter, EX4 4QL, United Kingdom;
$^{d}$Observatoire de la Côte d'Azur (France);
$^{e}$Sydney Institute for Astronomy, School of Physics, Physics Road, University of Sydney, NSW 2006, Australia;
$^{f}$Space sciences, Technologies \& Astrophysics Research (STAR) Institute, University of Li\`ege, Li\`ege, Belgium;
$^{g}$Institute for Particle Physics and Astrophysics, ETH Zurich, 8093 Zurich, Switzerland;
$^{h}$Department of Astronomy and Steward Observatory, 933 North Cherry Ave, Tucson, AZ 89 85721, USA;
$^{i}$Large Binocular Telescope Observatory, 933 North Cherry Ave, Tucson, AZ 85721, USA;
$^{j}$I. Physikalisches Institut, Universit\"at zu K\"oln, Z\"ulpicher Str. 77, 50937 Cologne, Germany;
$^{k}$European Organisation for Astronomical Research in the Southern Hemisphere, Casilla, 19001, Santiago 19, Chile;
$^{l}$Max-Planck-Institut für Astronomie, Königstuhl 17, 69117 Heidelberg, Germany.

\end{center}
~\newline

\begin{abstract}
ESO’s Very Large Telescope Interferometer has a history of record-breaking discoveries in astrophysics and significant advances in instrumentation. 
The next leap forward is its new visitor instrument, called Asgard. 
It comprises four natively collaborating instruments: HEIMDALLR, an instrument performing both fringe tracking and stellar interferometry simultaneously with the same optics, operating in the K band; Baldr, a Strehl optimizer in the H band; BIFROST, a spectroscopic combiner to study the formation processes and properties of stellar and planetary systems in the Y-J-H bands; and NOTT, a nulling interferometer dedicated to imaging nearby young planetary systems in the L band. 
The suite is in its integration phase in Europe and should be shipped to Paranal in 2025. 
In this article, we present details of the alignment and calibration unit, the observing modes, the integration plan, the software architecture, and the roadmap to completion of the project.
\end{abstract}


\section{Introduction}
\label{sec:intro}  
The emphatic triumph of the Very Large Telescope Interferometer (VLTI) and its second-generation instruments in delivering unique science has set European astronomy apart. 
The ongoing facility upgrade within the Gravity+ framework \citep{Eisenhauer2019} promises still further ground-breaking scientific discoveries. 
In parallel, major technology and scientific milestones have been achieved on other interferometric facilities, such as the Center for High Angular Resolution Astronomy Array (CHARA) and the Large Binocular Telescope Interferometer (LBTI). 
New ideas and laboratory demonstrations have also emerged in recent years opening the path to novel scientific capabilities for optical interferometry. 
Leveraging these recent developments, the Asgard instrument suite will extend the scientific capabilities of the VLTI with a set of four instrument modules: BIFROST \citep{bifrost_kraus, bifrost_chhabra, bifrost_mortimer} (Beam-combination Instrument for studying the Formation and fundamental paRameters of Stars and planeTary systems) which is a Y/J/H-band combiner optimized for high spectral resolution, Baldr which is an H-band injection optimizer for BIFROST, HEIMDALLR \citep{ireland2018, Taras2024} (High-Efficiency Multiaxial Do-it ALL Recombiner) which is a high-sensitivity K-band fringe tracker, and NOTT \citep{hi5_defrere, hi5_dandumontA, hi5_dandumontB, hi5_garreau, hi5_laugier, hi5_sanny} (Nulling Observations of dusT and planeTs) which is an L-band nuller optimized for high-contrast observations. 
Science cases and instrument capabilities have been extensively described in Martinod et al. (2023) \citep{Martinod2023JATIS}, Defrère et al. (2022) \citep{hi5_defrere} and Kraus et al. (2022) \citep{bifrost_kraus}.
This paper presents the source calibration unit, the observing modes, the integration plan, the software architecture, and the roadmap. 

\section{Observing modes}
\subsection{For NOTT}
\begin{figure}[h]
    \centering
    \includegraphics[width=0.6\textwidth]{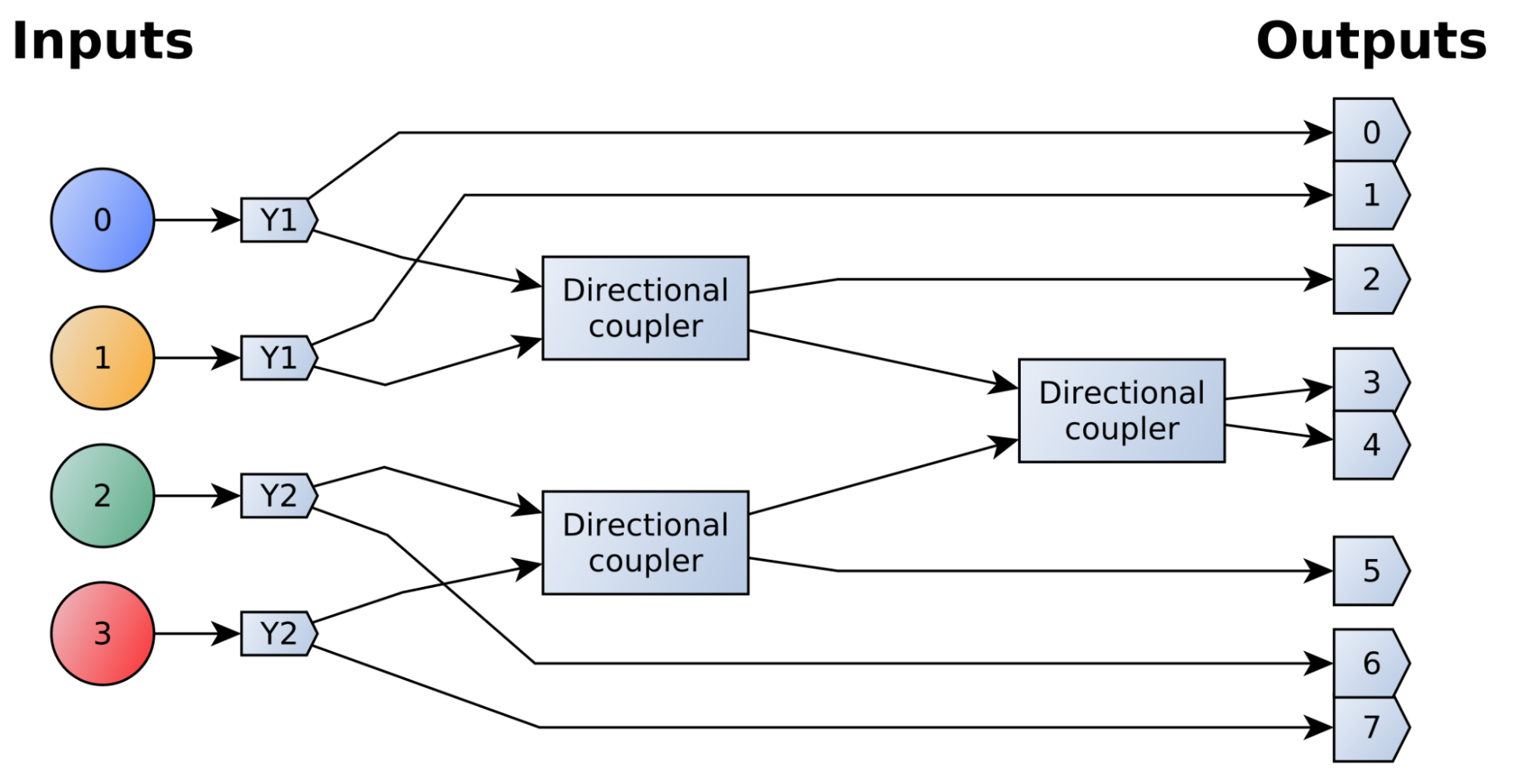}
    \caption{Diagram of the NOTT’s photonic beam combiner, which consists of photometric outputs and two stages of directional couplers. Outputs 0, 1, 6 and 7 are photometric (they give the intensities of the beams). Outputs 2 \& 5 collect the bright interferences of the first two combinations. Outputs 3 \& 4 collect the nulled signals in a Double-Bracewell mode.}
    \label{fig:nott_combiner}
\end{figure}

\begin{figure}[h]
    \centering
    \includegraphics[width=0.75\textwidth]{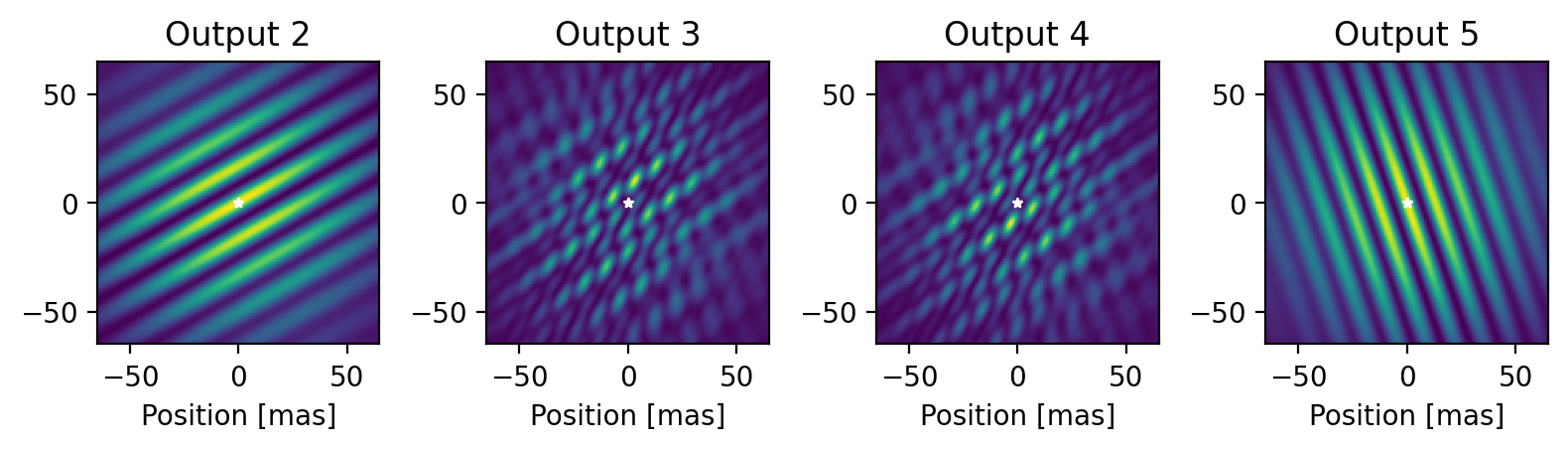}
    \caption{Transmission map of the interferometric outputs of NOTT’s photonic chip (see Fig.~\ref{fig:nott_combiner}) in the Planet mode. Outputs 2 and 5 provide simple-Bracewell transmission maps. Outputs 3 and 4 provide anti-symmetrics and complementary maps; subtracting them provides a quantity robust against leakage from a symmetric source.}
    \label{fig:nott_outputs_planet}
\end{figure}

NOTT implements the so-called Angel \& Woolf \citep{Angel1997} beam combination (or double Bracewell) scheme (Fig.~\ref{fig:nott_combiner}).
This combination allows three observing modes: (i) the planet mode, (ii) the disk mode and (iii) the simple-Bracewell mode.
The nulled outputs of two pair-wise 50/50 couplers are cross-combined in a second stage to a pair of complementary dark signals (Fig.~\ref{fig:nott_outputs_planet}). 
Using the difference of outputs 3 and 4 as primary observable offers robustness to instrumental errors. 

In planet mode, the phase-offset between the nulled input pairs $(0, 1)$ and $(2,3)$ is set to $\pi/2$ (Fig.\ref{fig:nott_combiner}). 
In this configuration, the differential output creates an anti-symmetric transmission map which isolates the planetary signal from any symmetric light such as the contributions of the star, thermal background, local zodiacal cloud, and instrumental biases \citep{Laugier2023}.

\begin{figure}[h]
    \centering
    \includegraphics[width=0.75\textwidth]{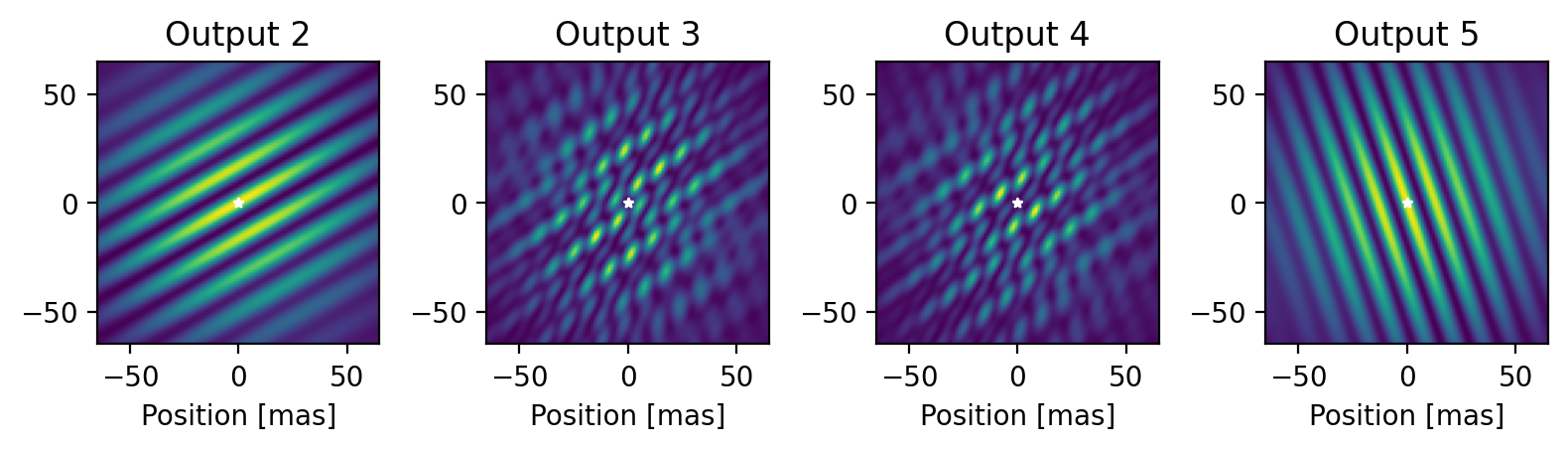}
    \caption{Transmission map of the interferometric outputs of NOTT’s photonic chip (see Fig.~\ref{fig:nott_combiner}) in the Disk mode. Outputs 2 and 5 provide simple-Bracewell transmission maps. Outputs 3 and 4 provide symmetric and complementary maps; substracting them provides a quantity robust against leakage from an anti-symmetric source.}
    \label{fig:nott_outputs_disk}
\end{figure}

In disk mode, the phase-offset between the two nulled pairs is set to $0$, which makes the transmission map of the differential output symmetric (Fig.~\ref{fig:nott_outputs_disk}). 
This makes NOTT exclusively sensitive to symmetrical features.

The simple-Bracewell mode is meant to be the first mode commissioned on sky, thanks to simpler implementation and reliance on preexisting data processing pipelines \citep{hanot2011, Mennesson2016, martinod2021}.
As opposed to both disk and planet modes, each of the first stage pairs are phased, producing two null outputs directly at outputs $2$ and $5$, making the nuller sensitive to symmetrical and anti-symmetrical features on two independent interferometric baselines. 
The constructive outputs are sent to the cross-combiner and can be discarded, while the photometric outputs are used to estimate the intensity of the constructive fringes of both baselines.

\subsection{For BIFROST}
In order to fulfill its scientific program \citep{bifrost_kraus}, BIFROST contains two spectrographs that can be configured separately: a low- (LR) and a high-resolution (HR) spectrographs, realised by two optical arms in the instrument.
The LR arm covers the spectral bands from Y to H while the HR arm only covers Y and J bands.
In addition to these spectroscopic capabilities, BIFROST is able to observe on- and off-axis, and to split the polarisation.
This flexibility allows use of BIFROST with 3 observing modes: (i) HIGHSENS, (ii) SPECTRO-YJ, and (iii) WOLL. 
A fourth mode (SPECTRO-H) is planned for and could be implemented subject to funding.

HIGHSENS is for high-sensitivity on-axis observations. 
Here, only the LR spectrograph is used. 
The spectrograph will observe in either Y-J bands, or H band.

SPECTRO-YJ  is the mode for which BIFROST is primarily built, with 90\% of the light going to the HR spectrograph and 10\% of the light going to the LR spectrograph. 
The LR spectrograph is configured for the Y and J bands and the HR spectrograph is configured with one of the high spectral resolution prisms/Volume Phase Holographic Gratings \citep{bifrost_chhabra} for spectral line observations. 
The telemetry from the LR spectrograph is reported back to Asgard, so that it can be used to track water vapour dispersion.
This effect induces non-linear phase shifts between the fringe tracker in the K-band and the nuller in the L-band.
Tracking this dispersion will help to improve the fringe tracking and the sensitivity of the nuller.
The telemetry is also used to optimise the Longitudinal Dispersion Correctors (LDC) settings. 
Later for data reduction, the LR spectrograph telemetry and flux measurements are used to apply a phasor correction to the HR spectrograph data and to reject frames with poor injection or fringe tracker drop-outs/phase wrapping.

SPECTRO-H gets light in the H-band spectral lines only.
This separation with the previous mode is motivated by optical constraints as VPHGs \citep{bifrost_chhabra} are used in the LR spectrograph. 
This mode only operates with the LR spectrograph and the on-axis mode.
The main reasons are the HR spectrograph is designed so that it is not sensitive in the H band to minimise thermal background and maximise the sensitivity.
Given the operation is reduced to a single spectrograph, it is not possible to apply the frame selection / phasor correction technique, but instead to rely on HEIMDALLR to provide stable fringe tracking. 
In a long-term future upgrade, H-band spectro-interferometry could be optimised by a small cryostat that would reduce thermal background for long exposures and enable the dual-spectrograph operation also for H-band.

WOLL (split-polarisation interferometry) measures the two polarisation states separately.
This is facilitated with a linear stage located behind the LR spectrograph filter wheel that can be used to move a Wollaston into the beam. 
This mode could be used for improving the visibility calibration. 
Together with half-wave plates (HWP), this mode could potentially also be used for observations in polarized light, although this is only a long-term goal and will require detailed modelling of the polarisation properties along the VLTI beam path. 
In this mode, the ON/OFF-axis beam splitter is moved out of the beam, allowing all light to proceed to the LR spectrograph. 

\section{Opto-mechanical design and calibration source unit}
\begin{figure}[h]
    \centering
    \begin{tabular}{cc}
         \includegraphics[width=0.48\textwidth]{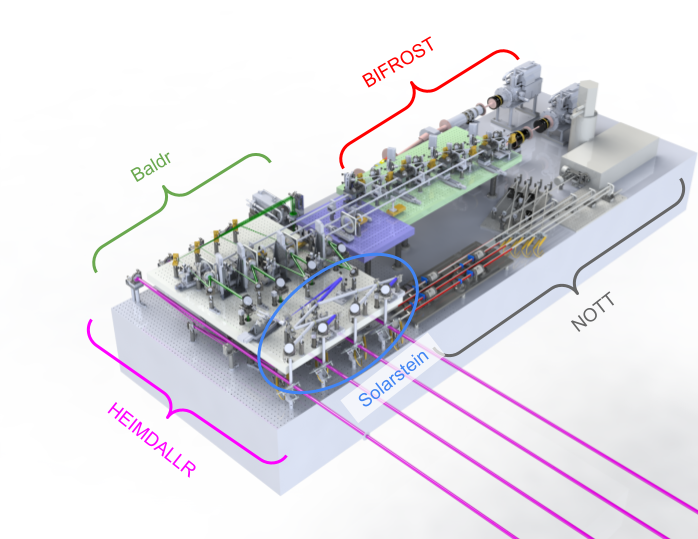} &
         \includegraphics[width=0.48\textwidth]{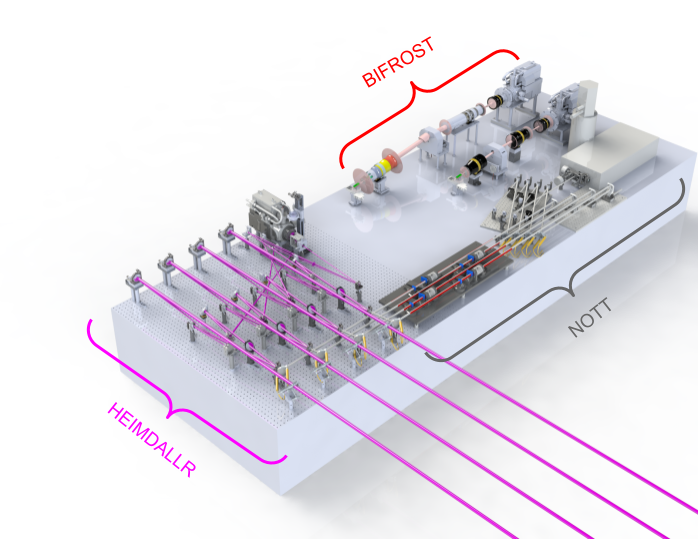} \\
    \end{tabular}
    \caption{Opto-mechanical design of Asgard. Left: The complete instrument suite, with the upper level shown. On this level, there are the optics of Baldr (Zernike wavefront sensor),  which shares a detector with Heimdallr. BIFROST (Y-H bands stellar interferometer) and Solarstein (calibration source unit generating calibration and alignment beams, utilizing injection mirrors to shift between the sky and internal source.) are also highlighted. Right: The lower level with the optics of HEIMDALLR (fringe tracker and stellar interferometer), BIFROST low- and high-resolution arms, and NOTT (L-band nulling interferometer) visible.}
    \label{fig:opto-mech}
\end{figure}

Asgard will be located on the Visitor 2 table, formerly the AMBER table. 
The opto-mechanical design of Asgard is given in Figure~\ref{fig:opto-mech}.
The layout is the most up-to-date but the descriptions of the components in previous publications are still valid \citep{Martinod2023JATIS, asgard_martinod}.

Here we describe the design of the calibration unit, called Solarstein \citep{Taras2024}.
This module contains several sources for alignment, spectral calibration and cophasing: 
\begin{itemize}
    \item a 532 nm laser (for alignment purposes)
    \item a 635 nm laser (for alignment purposes)
    \item a white thermal source (for cophasing purposes)
    \item a spectral source (for spectral calibration purposes)
\end{itemize}
The performances of the dichroics splitting the light in L, K, H-Y bands are not well specified in the visible.
Consequently, two visible lasers at two different wavelengths are selected to ensure enough flux to be transmitted for the alignment procedure.

\begin{figure}[h]
    \centering
    \includegraphics[width=0.5\textwidth]{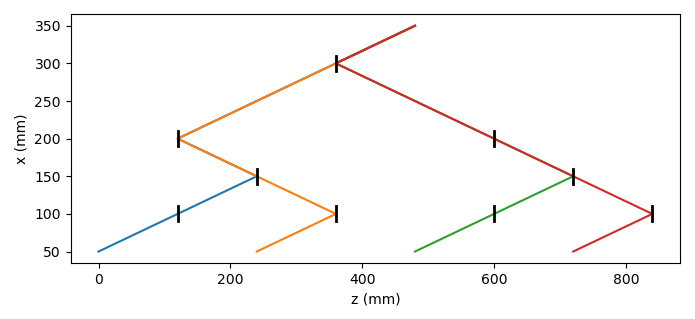}
    \caption{Design of the Solarstein beam maker, producing cophased and balanced beams. The beams are travelling from the top to the bottom of the figure. Three beam splitters are used to create four balanced beams. Plates are also added within some beam trains to ensure the equality of the optical path lengths thus the cophasing between the beams.}
    \label{fig:solarstein}
\end{figure}

A selection stage sends the light of one of these sources into a multimode optical fiber that leads to a pinhole to produced expanded beams which are collimated by off-axis parabolas in Solarstein \citep{Taras2024}.
The pinhole size is 20 microns, allowing only 1/400th of the white light to propagate. 
The light from the pinhole is then collimated by a 200~mm focal length off-axis parabola (OAP) and sent to an optical system comprised of 9 beam splitters to create 4 balanced and cophased beams (Fig.~\ref{fig:solarstein}).
Finally, flippable mirrors select between light from Solarstein and the VLTI.

\section{Control architecture}
\subsection{Overview}
\begin{figure}[h]
    \centering
    \includegraphics[width=0.9\textwidth]{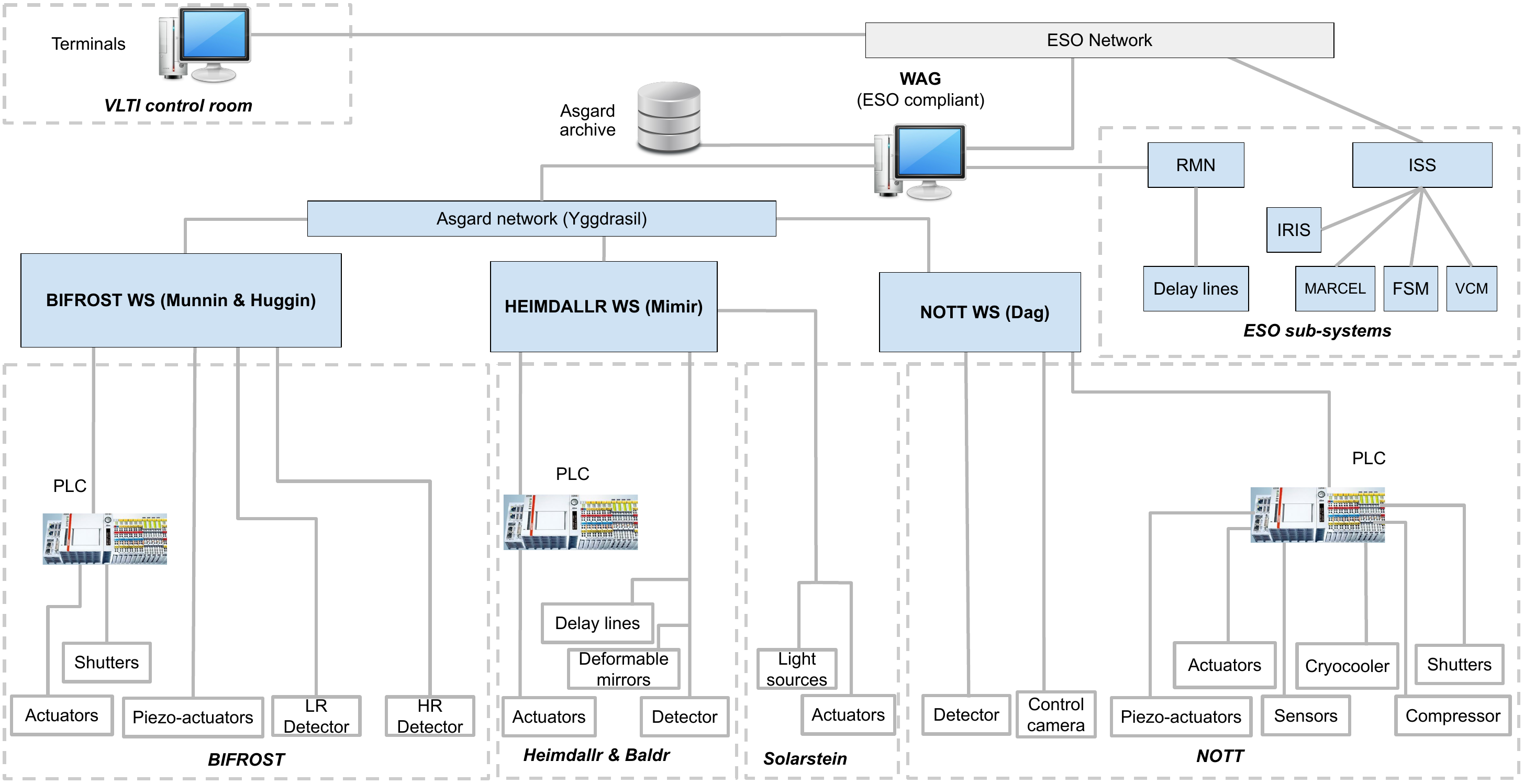}
    \caption{Control architecture of Asgard. WAG is an ESO-compliant machine that can communicate with ESO facility (in particular the RMN and the ISS) and control every Asgard subsystems through their respective workstations. Each instrument has one or two workstations to control its optics and cameras via programmable logic controllers (PLC). The users operate Asgard from terminals located in the VLTI control room that will connect to the WAG workstation.}
    \label{fig:asgard_architecture}
\end{figure}

The overall Asgard control architecture is given in Figure~\ref{fig:asgard_architecture}.
The hardware architecture is designed so that all the modules can be operated from the terminals in the VLTI control room.
This centralisation allows the observer to run all the instruments simultaneously and to communicate with the VLTI systems. 

Asgard low-level IT hardware and software are non-ESO compliant, hence must be kept separate. 
In addition, they need to communicate with each other; hence, they are connected to each other on a server-client system. 
Asgard modules need to interact with ESO systems such as the Reflective Memory Network (RMN), the delay lines and the calibration sources.
The solution is to connect all instruments’ workstations, via a local network called \emph{Yggdrasil}, to an ESO compliant machine, the instrument Workstation of AsGard, (WAG, formerly Hermod workstation) which is on ESO’s network. 
WAG will interact with ESO systems, the RMN, and transmit data to and from Asgard's modules that require such data. 
WAG respects ESO’s standards for hardware and software. 
From the ESO network point of view, Asgard is operated from a single computer represented by WAG.
Each instrument has one or two workstations called \emph{Module Control Units} (MCU):
\begin{itemize}
    \item The HEIMDALLR module control unit, named Mimir, reads out the image frames from HEIMDALLR \& Baldr's C-Red-1 camera, processes them in real-time and sends out Optical Path Difference (OPD) and wavefront correction data. Mimir directly controls the Deformable Mirrors (DMs) and the internal delay lines within HEIMDALLR;
    \item The NOTT module control unit, named Dag, controls the camera and the opto-mechanical devices of NOTT;
    \item The first BIFROST module control unit, named Munnin, controls the C-Red-1 low-spectral-resolution camera and records its data. It also controls the on-axis beam opto-mechanical devices of BIFROST;
    \item The second BIFROST module control unit, named Huginn, controls the C-Red-1 high-spectral-resolution camera and records its data. It also controls the off-axis beam opto-mechanical devices of BIFROST.
\end{itemize}

Besides controlling some sub-systems directly, the workstations are also using Programmable Logic Controllers (PLC).
The choice of driving a sub-system directly from the workstation or a PLC is motivated by the latency, frequency of usage and the most convenient way to drive the component while respecting constraints such as interface or power consumption.

Each instrument (except Solarstein) has its own data storage embedded in its respective MCU. 
Consistent with the requirement for VLTI Visitor Instruments to keep their data flow separate from that of ESO, Asgard also has a general archive, the Asgard Archive. 
The Asgard archive has sufficient capacity to store data acquired over several weeks. 
In a future upgrade, the Asgard archive could also serve as the hub for feeding Asgard data into the ESO archive. 
The workstations, which directly manage the detectors, have direct access to it to write the frames and the user can access the storage via WAG.

\subsection{Top-level control}
The Instrument Workstation WAG is the conductor of Asgard. 
It is a R750 Dell server running Linux. The Asgard top-level control software developed for WAG is based on the suite of tools, Tcl/Tk and C++ libraries known as the ``VLT-Software'', that is used in all the VLT and VLTI instruments \citep{Pozna2008}. 
The Asgard top-level control software on WAG bridges the gap between the Non-VLT-Software MCUs and the VLTI sub-systems. 
Providing a user interface common to VLT and VLTI instruments, it includes the following software components:
\begin{itemize}
    \item An Observation Software (OS) that receives commands from the Broker of Observing Blocks (BOB) and broadcasts them to the Instrument Control Software (ICS) and Detector Control Software (DCS) running on WAG, and to the Interferometric Supervisor Software. Following VLT standards, BOB is the main user interface of the instrument for scientific operations. The OS also includes special processes for the communications with the MCUs and with the RMN of the VLTI;
    
    \item An ICS that controls, through the MCUs, the motors of all the ASGARD modules, the lasers and lamps of SOLARSTEIN and the switch rack unit of BIFROST. The ICS also periodically reads the temperature and pressure sensors of NOTT's cryostat to monitor this instrument's health and safety. The deformable mirrors of BALDR are not part of the ICS and are controlled via HEIMDALLR's workstation. All the motors that are directly controlled by the ICS are actuated during VLTI preset (VLTI configurations and pointings done during an observing night) or between two detector exposures. They may take up to a few seconds to reach their selected positions. Some of them can also be directly controlled in real-time by the MCUs, using pre-defined trajectories (for atmospheric dispersion compensation) or results of detector data processing. 

    \item A DCS server for each detector (4 in total) in the ASGARD suite. These detectors do not use ESO-standard drivers and are controlled by the MCUs. Detector data are recorded on the MCUs. Therefore, the role of each DCS, on WAG, is to forward commands to its MCU counterpart, in order to set up the detector and to start or abort exposures. The DCS also allows to start and stop the real-time tasks running on the MCUs. Each DCS includes a Real-Time Display (RTD) interface that receives frame data and forwards them to the RTD server of WAG.
\end{itemize}

\subsection{Communication}
The physical layer for the communications between WAG and the MCUs is the YGGDRASIL internal network of ASGARD (Gigabit Ethernet). 
A dedicated port on WAG is used for the internal ASGARD communications on YGGDRASIL. 
WAG is also connected to the VLTI network through another port, in order to communicate with the VLTI workstation running the Interferometric Supervisor Software (ISS). 
Through the VLTI network, WAG will also access the VLT time server and broadcast a reference time to the MCUs through the NTP protocol.
The goal is to synchronize the actions of the different instruments to move from one configuration to another and the fringe tracking, wavefront control, acquisitions altogether. 
WAG is also connected to the RMN of the VLTI, in order to read the positions of the delay-lines and potentially control them (in a future upgrade of the Asgard suite). 
WAG can also read from the RMN the adaptive-optics status (wavefront errors).

\section{Integration and commissioning plan}
Asgard is a two-level instrument comprising 5 modules.
Such complexity involves an assembly, integration, verification and commissionning at Paranal in four phases.

\paragraph{Pre-phase 1:}
Assembly, Integration and Verification of HEIMDALLR, Baldr, NOTT and the injection module of BIFROST at Laboratoire Lagrange (Nice, France).
This stage will be validated by ESO in the first quarter of 2025. 
This will be followed by the shipment to Paranal and the integration of this first stage of Asgard at the VLTI.
BIFROST will be first integrated and subsequently validated by ESO at the University of Exeter before being shipped to Paranal later in 2025.

\paragraph{Phase 1:}
HEIMDALLR, Baldr, NOTT, and one injection module of BIFROST are assembled and integrated on the visitor 2 table at the VLTI. 
Commissioning of HEIMDALLR/Baldr performance on bright single targets.
NOTT will commission its simple-Bracewell mode first. 
Commissioning of BIFROST's injection module with Baldr will consist of evaluating how well the light from the Y to H bands is injected into the BIFROST fiber.
All the commissionings of this phase will be done with the 1.8 m Auxiliary Telescopes (ATs).
This phase will span two periods: one in May 2025 for daytime AIV only, and a second one with daytime AIV and one on-sky commissioning run in July 2025.

\paragraph{Phase 2:}
Assembly and integration of the BIFROST photonic chip and LR arm, which includes the low-resolution prism and medium-spectral resolution H-band gratings.
The on-axis mode of BIFROST \citep{bifrost_kraus} will be commissioned, as well as the faint modes for HEIMDALLR/Baldr and the dual-Bracewell mode of NOTT.
The fringe tracking and wavefront correction provided by Heimallr/Baldr will be refined, including resolved object fringe tracking and delay line interaction through the RMN.
All the commissionings of this phase will be done with the ATs.
The phase 2 will start no earlier than November 2025.

\paragraph{Phase 3:}
The BIFROST High-Resolution arm and off-axis mode will be added. 
BIFROST off-axis \citep{bifrost_kraus} will be commissioned on the ATs and all the Asgard observation modes mentioned above will be commissioned on the 8 m Unit Telescopes (UTs).
An additional Daytime AIV of 6 days will be required.
This phase is expected to begin around May 2026.

\section{Conclusion}
Asgard, the new visitor instrument of the VLTI will open new unique capabilities (Y-J-H high-spectral resolution and L-band high-contrast nulling interferometry) and enable exoplanet imaging at milli-arcsecond angular separation.
It consists of 4 different modules: BIFROST (YJH high-spectral resolution beam combiner), Baldr (H band Zernike Wavefront Sensor),  HEIMDALLR (dual K band high-sensitivity fringe tracker), and NOTT (L-band nulling interferometer based on a photonic chip and self-calibration data reduction techniques).

BIFROST has four observing modes (Highsens, Spectro-YJ, Spectro-H and Woll) and NOTT has three (planet, disk and simple-Bracewell).
The opto-mechanical design is mostly final as well as the overall structure of the control scheme.
The latter consists of a conductor machine called WAG, being the interface between ESO and Asgard sub-systems and 4 workstations to drive the sub-systems (1 for HEIMDALLR and Baldr, 1 for NOTT and 2 for BIFROST).
The AIV \& commissioning in VLTI follows a three-tiered approach.

\section*{Acknowledgments}
M-A.M. has received funding from the European Union’s Horizon 2020 research and innovation programme under grant agreement No.\ 101004719.
SK acknowledges support from STFC Consolidated Grant (ST/V000721/1).

D.D. acknowledges support from the European Research Council (ERC) under the European Union's Horizon 2020 research and innovation program (grant agreement No.\ CoG - 866070).

S.K., S.C., J.P., and O.S. acknowledge support from an ERC Consolidator Grant (``GAIA-BIFROST'', grant agreement No.\ 101003096). 

A.T., P.T., J.B., F.C. and B.N. acknowledge support from Astralis - Australia's optical astronomy instrumentation Consortium - through the Australian Government’s National Collaborative Research Infrastructure Strategy (NCRIS) Program as well as an Australian Research Council (ARC) Linkage Infrastructure Funding (LIEF) grant LE220100126. 
This work used the ACT, SA and Sydney nodes of the NCRIS-enabled Australian National Fabrication Facility.

D.M. and O.S. acknowledge support from STFC Consolidated Grant (ST/V000721/1).

S.E. is supported by the National Aeronautics and Space Administration through the Astrophysics Decadal Survey Precursor Science program (Grant No. 80NSSC23K1473).

\bibliography{report} 
\bibliographystyle{plainnat}

\end{document}